\documentclass[aps,epsf,rotate,nofootinbib,citesort,preprint]{revtex4}

\usepackage{graphicx}
\usepackage{times}
\usepackage{natbib}

\begin{document}

\title{Log-normal statistics in e-mail communication patterns}

\author{Daniel B.~Stouffer}
\thanks{These two authors contributed equally to this work.}
\affiliation{\scriptsize 
Department of Chemical and Biological Engineering,
Northwestern University, Evanston, IL 60208, USA}

\author{R.~Dean Malmgren$^{\ast}$}
\affiliation{\scriptsize
Department of Chemical and Biological Engineering,
Northwestern University, Evanston, IL 60208, USA}

\author{Lu\'is A.~N.~Amaral}
\email{amaral@northwestern.edu}
\affiliation{\scriptsize
Department of Chemical and Biological Engineering,
Northwestern University, Evanston, IL 60208, USA}

\date{\today}

\begin{abstract}
  Following up on Barab\'asi's recent letter to \textit{Nature} [{\bf
  435}, 207--211 (2005)], we systematically investigate the time
  series of e-mail usage for 3,188 users at a university.  We focus on
  two quantities for each user: the time interval between
  consecutively sent e-mails (interevent time), and the time interval
  between when a user sends an e-mail and when a recipient sends an
  e-mail back to the original sender (waiting time).
  We perform a standard Bayesian model selection analysis that
  demonstrates that the interevent times are well-described by a
  single log-normal while the waiting times are better described by
  the superposition of two log-normals.  Our analysis rejects the
  possibility that either measure could be described by truncated
  power-law distributions with exponent $\alpha \simeq 1$.
  We also critically evaluate the priority queuing model proposed by
  Barab\'asi to describe the distribution of the waiting times.  We show
  that neither the assumptions nor the predictions of the model are
  plausible, and conclude that a theoretical description of human
  e-mail communication patterns remains an open problem.
\end{abstract}

\maketitle

\section{Introduction} \label{sect:intro}

Human beings are extraordinarily complex agents.  Remarkably, in spite
of that complexity a number of striking statistical regularities are
known to describe individual and societal human behavior
\cite{stanley96,amaral97a,amaral98,plerou99,amaral01,guimera02}.
These regularities are of enormous practical importance because of the
influence of individual behaviors on social and economic outcomes.

Even though the analysis of social and economic data has a long and
illustrious history, from Smith~\cite{smith86} to
Pareto~\cite{pareto06} and to Zipf~\cite{zipf49}, the recent
availability of digital records has made it much easier for
researchers to \textit{quantitatively} investigate various aspects of
human behavior. In particular, the availability and omnipresence of
e-mail communication records is attracting much attention
\cite{ebel02,guimera03,eckmann04,barabasi05,watts06}.

Recently, Barab\'asi studied the e-mail records of users at a
university and reported two patterns in e-mail
communication~\cite{eckmann04}: the time interval between two
consecutive e-mails sent by the same user, which we will denote as the
interevent time $\tau$, and the time interval between when a user
sends an e-mail and when a recipient sends an e-mail back to the
original sender, which we will denote as the waiting time $\tau_w$,
follow power-law distributions which decay in the tail with exponent
$\alpha \simeq 1$.  Additionally, Barab\'asi proposed a priority
queuing model that reportedly captures the processes by which
individuals reply to e-mails, thereby predicting the probability
distribution of $\tau_w$.

Here, we demonstrate that the empirical results reported in
Ref.~\cite{barabasi05} are an artifact of the data analysis.  We
perform a standard Bayesian model selection analysis that demonstrates
that the interevent times are well-described by a single log-normal
while the waiting times are better described by the superposition of
two log-normals.  Our analysis rejects beyond any doubt the
possibility that the data could be described by truncated power-law
distributions.

We also critically evaluate the priority queuing model proposed by
Barab\'asi to describe the observed waiting time distributions.  We
show that neither the assumptions nor the predictions of the model are
plausible. We thus conclude that the description of human e-mail
communication patterns remains an open problem.

The remainder of this paper is organized as follows.  In Section II,
we describe the preprocessing of the data.  We then analyze the
distribution of interevent times (Section \ref{sect:tau}) and the
distribution of waiting times (Section \ref{sect:tauw}).  Finally, in
Section~\ref{sect:pqm} we investigate the priority queuing model
of Ref.~\cite{barabasi05}.

\section{Preprocessing of the data} \label{sect:data}

%
We consider here the database investigated by
Barab\'asi~\cite{barabasi05}, which was also the focus of an earlier
paper by Eckmann et al.~\cite{eckmann04}.  This database consists of
e-mail records for 3,188 e-mail accounts at a university covering an
83-day period.  Each record comprises a sender identifier, a recipient
identifier, the size of the e-mail, and a time stamp with a precision
of one second.  Before describing our analysis of the data, we first
note some important features of the data which impact the analysis.

The first important fact is that the data were gathered at an e-mail
server, not from the e-mail clients of the individual users.  It is
quite possible that some users have e-mail clients, like Microsoft
Outlook, which permit users to send multiple e-mails at once
regardless of when the e-mails were composed.  Moreover,
servers may parse long recipient lists into several shorter lists
\cite{berson92}.  For this reason, e-mails to multiple recipients were
occasionally recorded in the server as \textit{multiple} e-mails.
Each of these duplicate e-mails was then sent in rapid succession to a
different subset of the list of recipients in the actual e-mail.  Both
the client-side and server-side uncertainties introduce artifacts in
the time series of interevent times for each user as it could appear
that a user is sending several e-mails over a very short time
interval.

To minimize these uncertainties, we preprocessed the data in order to
focus on \textit{actual} human behavior.  First, we identify sets of
e-mails sent by a user that have the \emph{exact} same size but whose
time stamp differs by at most five seconds\footnote{Five seconds
corresponds with the average minimal bound on humanly possible
interevent times based on the experiment in Fig.~\ref{fig2a}.}.  We
then remove all but the first e-mail from the time series of e-mails
sent, while adjusting the list of recipients to the first e-mail to
include all recipients in the removed e-mails
\footnote{A more aggressive preprocessing method would also remove
  blind-carbon-copied (BCC) e-mails.  The basic idea is that an e-mail
  with BCC recipients will have its size increased by a few bytes due
  to the addition of outgoing headers.  E-mails to the BCC recipients
  would be sent by the server shortly after the e-mail to visible
  recipients and would thus increase the number of very small
  interevent times.  We choose to err on the side of caution and
  \emph{not} attempt to remove e-mails with BCC recipients, as their
  detection is more subjective.}.

\begin{figure}
\centerline{\includegraphics*[width=0.8\columnwidth]{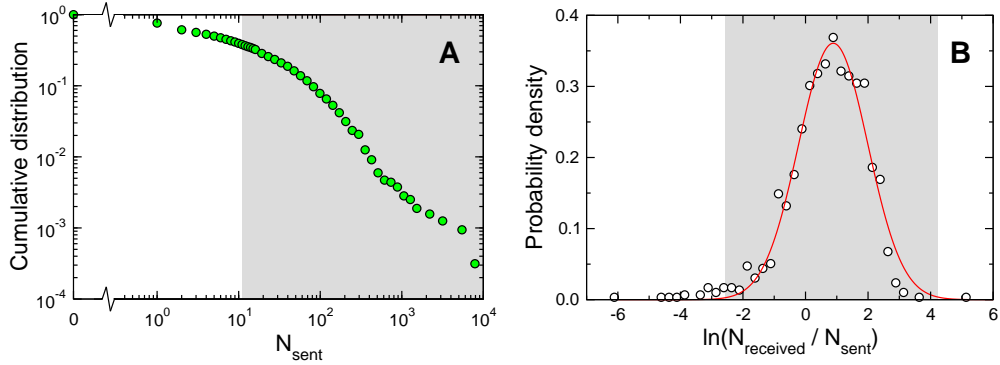}}
\vspace*{-0.3cm}
\renewcommand{\baselinestretch}{1.0}
\caption{Preprocessing of the data.  {\bf A}, The cumulative
  distribution of the number of e-mails sent by the 3,188 users over
  83 days.  To avoid characterizing users which use e-mail
  infrequently, we consider the 1,212 users which sent at least 11
  e-mails over 83 days (shaded region).  Note that this removes the
  759 users which sent no e-mails and the 370 users that sent one
  e-mail.  {\bf B}, The distribution of the ratio of the number of
  e-mails received to the number of e-mails sent is well-described by
  a log-normal (red line).  We use this fact to develop a criteria to
  identify \emph{typical} e-mail users that sent at least 11 e-mails
  over 83 days and differentiate them from bulk e-mail accounts,
  listserves, and e-mail accounts that are rarely used by their
  owners.  We keep the 1,152 users which fall within three standard
  deviations of the mean (shaded region). }
\label{parsing}
\end{figure}

An additional important fact to note is that some of the e-mail
accounts do not belong to ``typical'' users.  For example, User 1962
only sent 5 e-mails while receiving 2,284 e-mails.  This individual's
e-mail use is too infrequent to provide useful information on human
dynamics.  Meanwhile User 4099 sent 9,431 e-mails while receiving no
e-mails.  Although it cannot be confirmed due to the anonymous nature
of the data, this e-mail account was in all likelihood used for bulk
e-mails, implying that it cannot provide information on {\it human\/}
e-mail usage.

To avoid having our analysis distorted, we first restrict our
attention to users which sent at least 11 e-mails over the 83-day
experiment, yielding a minimum of 10 interevent times.  Our reasoning
is that users sending fewer e-mails do not use e-mail regularly enough
to allow us to truly infer patterns of human dynamics.  This procedure
excludes 1,976 of the 3,188 original e-mail accounts.

Next we examine the ratio of the number of e-mails received to the
number of e-mails sent to determine what constitutes a ``typical''
user.  This ratio is well-described by a log-normal distribution, and
we use this fact to consider only those users in our study who are
within three standard deviations from the mean.  This added constraint
excludes an additional 46 users.  We thus focus here on the 1,152
users who fulfill the above criteria (Fig.~\ref{parsing}).

\section{Interevent times} \label{sect:tau}

Reference \cite{barabasi05} reports that the probability distribution of
time intervals $\tau$ between consecutive e-mails sent by an
individual follows a power-law $P(\tau)\approx\tau^{-\alpha}$ with
$\alpha\simeq1$.  A basic examination of Barab\'asi's results,
however, quickly reveals a number of issues.

\begin{figure}
\centerline{\includegraphics*[width=0.8\columnwidth]{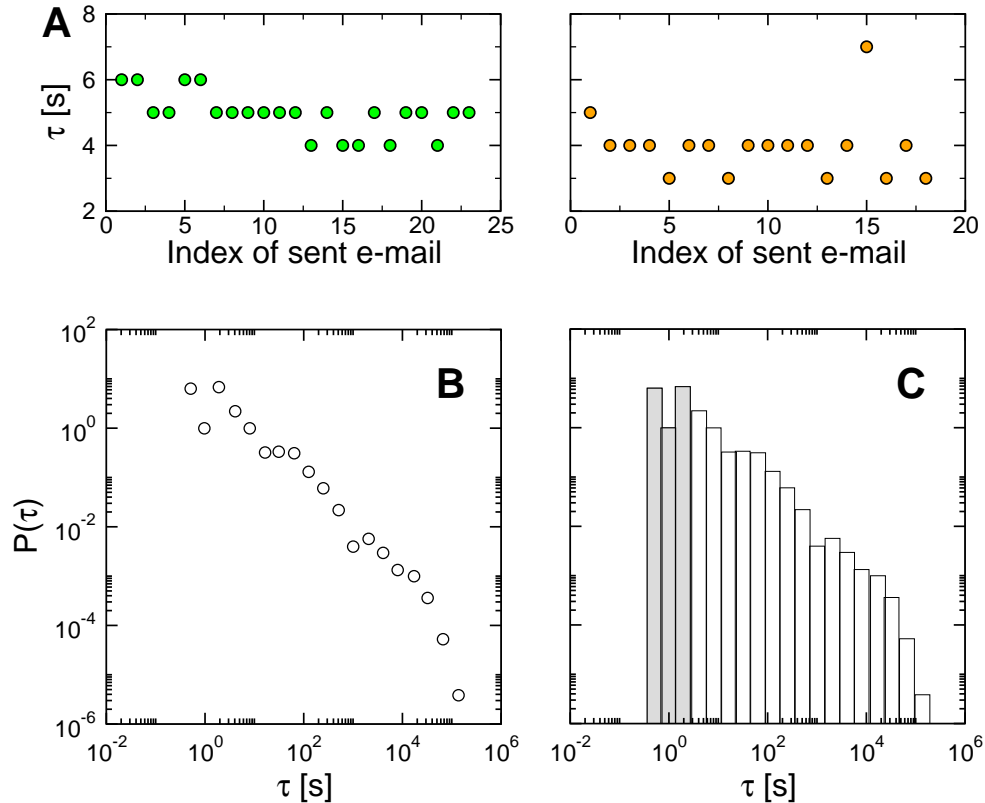}}
\vspace*{-0.3cm}
\renewcommand{\baselinestretch}{1.0}
\caption{The statistical analysis presented in Fig.~2a of
  Ref.~\cite{barabasi05}.  {\bf A}, Estimation of a lower bound on
  interevent times, $\tau$. Two of us sent about twenty e-mails trying to
  minimize the time interval between consecutive e-mails.  To be as
  fast as possible, we sent replies to an e-mail already in our
  inboxes.  Additionally, we did not even write any text or read the
  e-mail to which we were responding.  We find that humans need at
  least 3 seconds to send consecutive e-mails. {\bf B}, Reproduction of
  Fig.2a of Ref.~\cite{barabasi05} obtained with
  VistaMetrix~\cite{skillcrest04} and, {\bf C}, the same data with the
  boundaries of the bins clearly marked.  We assumed that the data
  points in Fig.~2a of Ref.~\cite{barabasi05} were placed in the
  middle of the bin. Note that there is a bin recording data for
  $\tau<1$ second, whereas the data have a resolution of one
  second.  The shaded bins indicate values with $\tau \le 3$ second,
  which contain 9\% of all events for the unidentified user.  }
\label{fig2a}
\end{figure}

\begin{enumerate}
\item Figure 2a of Ref.~\cite{barabasi05} features three bins
corresponding to interevent times $\tau \le 3$ seconds, an unphysical
interval (Fig.~\ref{fig2a}A).  The events in those bins in fact account
for 9\% of all events.

\item Figure 2a of Ref.~\cite{barabasi05} features at least one bin
confined to interevent times $\tau<1$ second (Fig.~\ref{fig2a}B--C)
while the data have a precision of one second~\cite{eckmann04}.
\end{enumerate}

We next quantitatively compare the plausibility of our log-normal
hypothesis with the plausibility of the power-law hypothesis of
Ref.~\cite{barabasi05} for interevent times.  To simplify the
analysis, we do not consider $\tau$, but its logarithm.  If a random
variable $\tau$ is log-normally distributed, then $u=\ln(\tau)$
follows a Gaussian distribution, whereas if $\tau$ is distributed
according to a power-law with exponent $\alpha = 1$, then $u$ is
uniformly distributed in the interval $\left[ \ln(\tau_{{\rm min}}),
\ln(\tau_{{\rm max}})\right]$.  Specifically, for
\begin{equation}
P(\tau) \propto \left\{
\begin{array}{ll}
\tau^{-1}\,, & \tau_{\rm min} \le \tau \le \tau_{\rm max} \\
0\,, & {\rm otherwise}
\end{array}
\right.
\label{powerlawpdf}
\end{equation}
the distribution of $u=\ln(\tau)$ is 
\begin{equation}
P(u) = \left\{
\begin{array}{ll}
\frac{1}{u_{\rm max} - u_{\rm min}}\,, & u_{\rm min} \le u \le u_{\rm
max} \\ 
0\,, & {\rm otherwise}
\end{array}
\right.
\label{uniformpdf}
\end{equation}
%

\begin{figure}
\centerline{\includegraphics*[width=0.8\columnwidth]{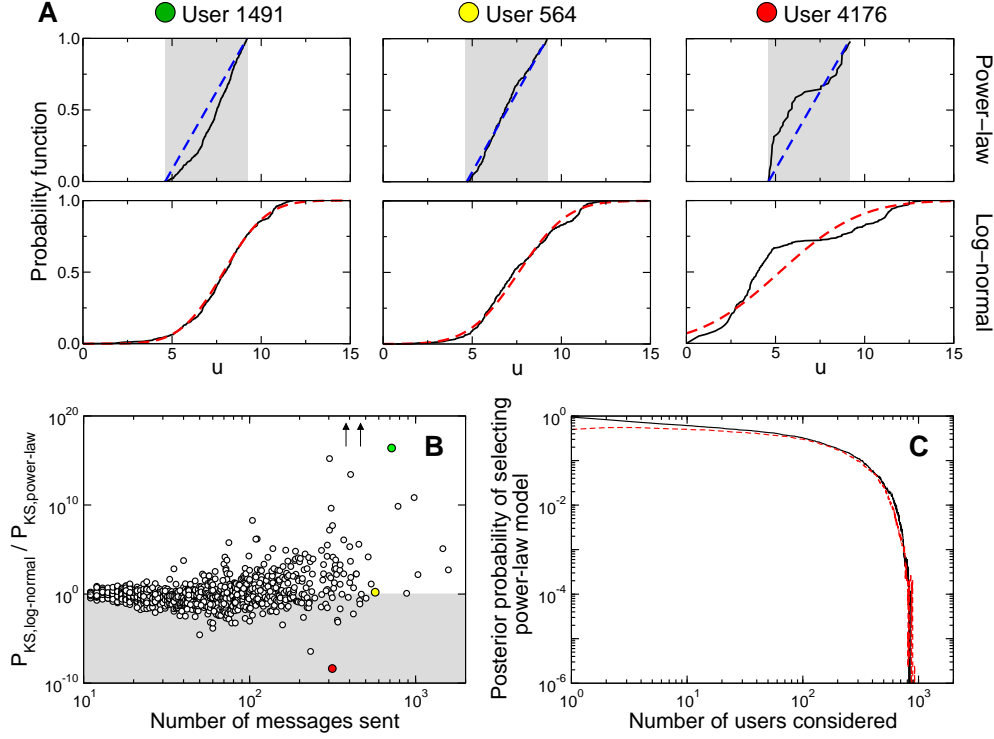}}
\vspace*{-0.3cm}
\renewcommand{\baselinestretch}{1.0}
\caption{Bayesian model selection protocol for interevent times.  {\bf A},
  Cumulative distributions of $u=\ln(\tau)$ for three users in the
  database.  The top panels show data and Gaussian model predictions
  for the entire range of $u$, whereas the bottom panels show data and
  power-law model for intermediate values of $u$.  {\bf B}, Scatter plot of
  $P_{\rm KS,log-normal}/P_{\rm KS,power-law}$, the ratio of the two
  $P_{\rm KS}$ values, for all available users depending on the number
  of consecutive e-mails sent for each user.  The larger circles
  colored green, yellow, and red correspond to the data shown in (A).
  Note that there are 2 users for which the ratio of $P_{\rm KS}$ is
  greater than $10^{20}$. Those users are indicated by the arrows. {\bf C},
  Recursively calculated posterior probability of accepting the
  power-law model. We use Bayesian model selection to recursively
  calculate the posterior probability of selecting the power-law
  distribution for two different prior probabilities: $P({\rm
  power-law})=0.95$ (solid black) and $P({\rm power-law})=0.50$ (red
  dashed).  In both cases, the posterior probability of selecting the
  power-law model {\it vanishes\/} after considering 932 of the 1,016
  users meeting our preprocessing criteria.  }
\label{ks_sent}
\end{figure}

Barab\'asi \cite{barabasi05} has argued that the power-law model is
meant to describe only ``intermediate'' $\tau$ values falling between
100 and 10,000 seconds.  Since some users have a smaller range of
$\tau$ values than that interval, we test the agreement of the
predictions of the power-law model only with data in the interval
$[\tau_{\rm min},\tau_{\rm max}]$, where $\tau_{\rm min} = \inf
\left\{ \tau | \tau \ge 100 \right\}$ and $\tau_{\rm max} = \sup
\left\{ \tau | \tau \le {\rm 10,000} \right\}$.  To properly specify
the power-law distribution, we must have at least two data points in
[$\tau_{\rm min},\tau_{\rm max}]$.  This constraint leads
to the exclusion of an additional 136 users.

We then use the Kolmogorov-Smirnov (KS) test~\cite{mood74} as a
measure of the plausibility of a model given the user's data.
Specifically, we compare the distribution of the logarithm of the
interevent times for a given user to two candidate models: a
Gaussian distribution and a uniform distribution.

Importantly, \textit{there is absolutely no fitting in our analysis}.
The parameters of the Gaussian distribution, $\mu$ and $\sigma$, are
simply the sample average and standard deviation of $u$, while the uniform
distribution is completely specified by $u_{\rm min}$ and $u_{\rm
max}$.  Figure \ref{ks_sent}D displays the ratio of the two KS
probabilities versus number of e-mails sent for all users with at
least two data points in the interval $[\tau_{{\rm
min}},\tau_{{\rm max}}]$.

In order to determine which of the two models provides a more accurate
description of the empirical data, we use the results of the KS test
as inputs in a Bayesian model selection
analysis~\cite{mood74,bernardo00}.  Bayes' rule states that
\begin{equation}
P\left(\mathcal{M}_j|\mathcal{E}_i\right) = \frac{P\left(\mathcal{E}_i|\mathcal{M}_j\right)
P(\mathcal{M}_j)}{\sum_{k} P\left(\mathcal{E}_i|\mathcal{M}_k\right) P(\mathcal{M}_k)}\,,
\label{bayes}
\end{equation}
where $P(\mathcal{M}_j|\mathcal{E}_i)$ is the posterior probability of
selecting model $\mathcal{M}_j$ given an observation $\mathcal{E}_i$,
$P(\mathcal{E}_i|\mathcal{M}_j) = P_{KS}(\mathcal{E}_i |
\mathcal{M}_j)$ is the probability of observing $\mathcal{E}_i$ given
a model $\mathcal{M}_j$, and $P(\mathcal{M}_j)$ is the prior
probability of selecting model $\mathcal{M}_j$.  Assuming no prior
knowledge about the correctness of the power-law and log-normal
models, one would select $P($log-normal$)=P($power-law$)=0.5$ for each
model.  However, to eliminate any bias on our part, we perform the
Bayesian model selection analysis for two cases: (\emph{i}) no prior
knowledge, $P($log-normal$)=P($power-law$)=0.5$, and (\emph{ii}) the
power-law model is far more likely to be correct,
$P($power-law$)=0.95$.

The availability of data for multiple users enables us to perform this
analysis recursively to obtain posterior probabilities of selecting
each model given the available data.  Concretely, the analysis of the
interevent times $\mathcal{E}_i$ from user $i$ updates the posterior
probabilities of the two models $P(\mathcal{M}_j|\mathcal{E}_i)$ using
Eq.~(\ref{bayes}).  These updated posterior probabilities are then
used as prior probabilities for the next user $i+1$. When all of the
users have been included, this analysis reveals the posterior
probability of the model given all of the available data.  The
Bayesian model selection analysis demonstrates that the likelihood of
the truncated power-law model being a good description of the data
vanishes to zero when all data is considered (Fig.~\ref{ks_sent}C).

\section{Waiting times} \label{sect:tauw}

Before we present our analysis of the waiting times, we must note that
the database collected by Eckmann et al.~\cite{eckmann04} and
analyzed by Barab\'asi \cite{barabasi05} is not particularly
well-suited for identifying the waiting times for replying to an
e-mail.  The data merely records that an e-mail was sent by user A to
user B at time $t$.  The data does not specify whether the e-mail from
A to B is, in fact, a reply to a prior message.  Imagine the following
scenario: user A sends an e-mail to user B.  Three days later, user B
sends an unrelated e-mail to user A.  Barab\'asi's approach
\cite{barabasi05}, which we follow, is to classify this e-mail as a
reply to the e-mail sent by user A three days earlier.  As this case
illustrates, {\it the analysis of waiting times is significantly less
reliable than that of interevent times.}

\begin{figure}
\centerline{\includegraphics*[width=0.8\columnwidth]{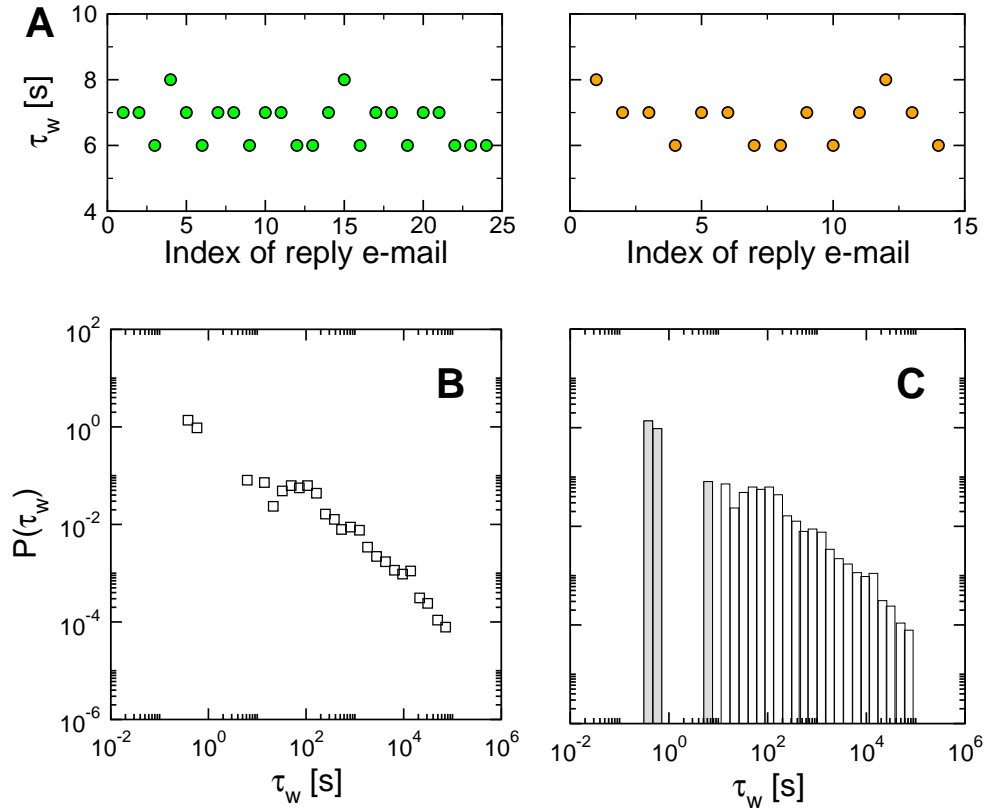}}
\vspace*{-0.3cm}
\renewcommand{\baselinestretch}{1.0}
\caption{The statistical analysis presented in Fig.~2b of
  Ref.~\cite{barabasi05}.  {\bf A}, Estimation of a lower bound on
  waiting time, $\tau_w$. Two of us sent about twenty replies to an
  e-mail already in our inboxes. To minimize the time required to do
  this, we did not read the e-mail to which we were responding but
  simply wrote ``yes'' at the top of our reply and then clicked send.
  We find that 6 seconds is the smallest waiting time feasible for a
  human.  {\bf B}, Reproduction of Fig.2a of Ref.~\cite{barabasi05}
  obtained with VistaMetrix~\cite{skillcrest04} and, {\bf C}, the same
  data but with the boundaries of the bins clearly marked.  We assumed that the data points in
  (B) are placed in the middle of the bin. Note that there are two bins
  recording data for $\tau_w<1$ second, while the data have a
  resolution of one second.  The shaded bins indicate values with
  $\tau_w < 6$ seconds consisting of 1\% of the data.  }
\label{fig2b}
\end{figure}

Reference \cite{barabasi05} reports that the probability distribution
of time intervals $\tau_w$ between receiving a message from a sender
and sending another e-mail to that sender follows a power-law
distribution $P(\tau_w)\approx\tau_w^{-\alpha}$ with $\alpha\simeq1$.
A cursory analysis of this result again reveals several problems.
\begin{enumerate}
\item Figure 2b of Ref.~\cite{barabasi05} features three bins
  corresponding to waiting times $\tau_w \le 6$ seconds, an unphysical
  interval (Fig.~\ref{fig2b}A).  The events in those bins account for
  1\% of all events.  

\item Figure 2b of Ref.~\cite{barabasi05} features two bins confined
  to waiting times $\tau_w < 1$ second (Fig.~\ref{fig2b}B--C) while the
  data have a precision of one second~\cite{eckmann04}.
\end{enumerate}

\begin{figure}
\centerline{\includegraphics*[width=0.8\columnwidth]{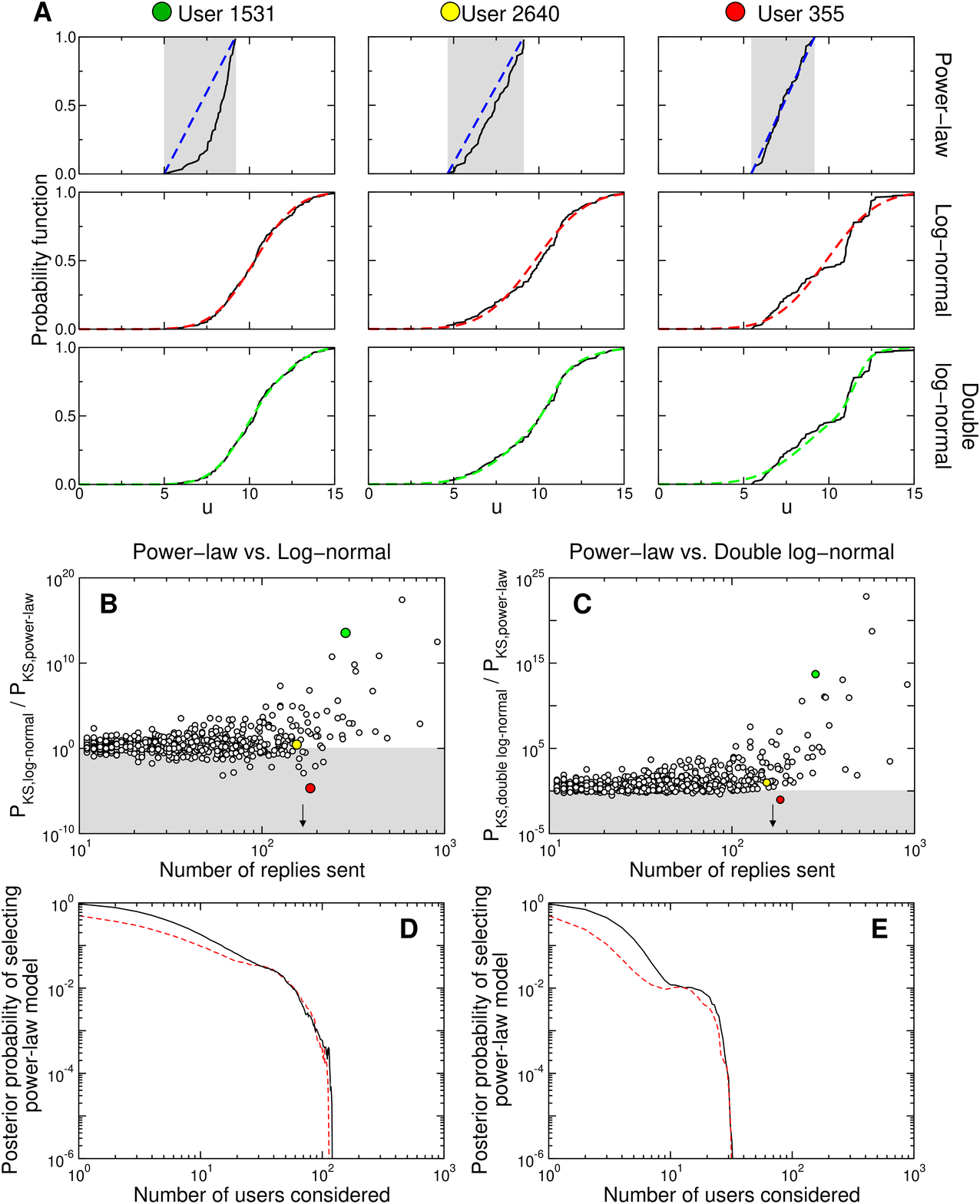}}
\vspace*{-0.3cm}
\renewcommand{\baselinestretch}{1.0}
\caption{Bayesian model selection protocol for waiting times.  {\bf A},
  Cumulative distributions of $u=\ln(\tau)$ for three users in the
  database.  The top panels show data and power-law predictions over
  intermediate values of $u$ whereas the middle and bottom panels
  depict Gaussian and double Gaussian model predictions for the entire
  range of $u$.  {\bf B}--{\bf C}, Scatter plot of the ratio of the
  two $P_{\rm KS}$ values for all available users depending on the
  number of e-mails sent for each user.  The larger circles
  highlighted in green, yellow, and red correspond to the data shown
  in (A).  Users outside the domain are indicated by the arrows.
  {\bf D}--{\bf E}, Recursively calculated posterior probability of
  accepting the power-law model for in comparison with the log-normal
  and double log-normal models. We use Bayesian model selection to
  recursively calculate the posterior probability of selecting the
  power-law distribution for two different prior probabilities:
  $P({\rm power-law})=0.95$ (solid black) and $P({\rm
  power-law})=0.50$ (red dashed).  The posterior probability of
  selecting the power-law model vanishes after considering 140 and 49
  of the 724 users for the log-normal and double log-normal
  comparisons, respectively.  }
\label{ks_waiting}
\end{figure}

We characterize the actual distribution of waiting times $\tau_w$
following the same procedure outlined in Section~\ref{sect:tau}.
After parsing the data, we are left with 724 users which have sent at
least 10 response e-mails over 83 days and have at least two waiting
times in the interval $100 \le \tau_w \le 10,000$ seconds.  We then
perform KS tests and Bayesian model selection to determine whether the
waiting times are better described by a power-law or log-normal
distribution.  The Bayesian model selection analysis demonstrates that
the likelihood of the truncated power-law model being a good
description of the data vanishes to zero when all data is considered
(Fig.~\ref{ks_waiting}).

\subsection{Double log-normal description}

Analysis of the data for the users with the largest number of replies
suggests that $\tau_w$ may actually be better described by a
superposition of two log-normal peaks: the first peak---which contains
most of the probability mass---typically corresponds with waiting
times of an hour, and the second peak typically corresponds with
waiting times of two days.  This finding prompted us to investigate
whether the superposition of two log-normals would provide a better
description of the data than a single log-normal.  The probability
function in this case has the functional form:
\begin{equation}
  F(u_w) = 0.5 \left[1 + f~{\rm erf}\left({\frac{u_w-\mu_{1}}{\sigma_{1}\sqrt{2}}}
    \right) + (1-f)~{\rm erf}\left({\frac{u_w-\mu_{2}}{\sigma_{2}\sqrt{2}}}
    \right)\right]\,,
\label{eqn:dln}
\end{equation}
where $\mu_1$ and $\mu_2$ are the means the two peaks, $\sigma_1$ and
$\sigma_2$ are the standard deviations of the two peaks, and $f$ is
the probability mass in the first peak.

\begin{figure}
\centerline{\includegraphics*[width=0.4\columnwidth]{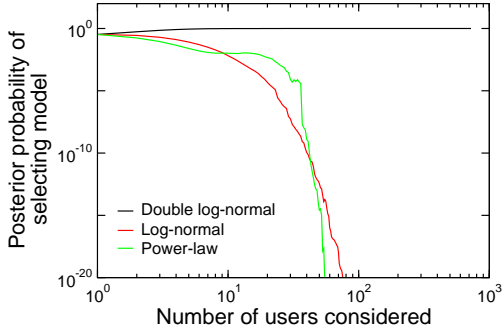}}
\vspace*{-0.3cm}
\renewcommand{\baselinestretch}{1.0}
\caption{Bayesian model selection protocol for comparing the double
  log-normal, log-normal, and power-law with exponent $\alpha=1$
  distributions.  After considering all of the 724 available users,
  the posterior probability of the power-law and log-normal vanishes.
  }
\label{threeway}
\end{figure}

In order to conduct the KS tests and Bayesian model selection, we must
first estimate the parameters of the double log-normal distribution,
Eq.~(\ref{eqn:dln}).  Unlike the earlier analyses, it is not possible
to estimate the parameters of the distribution without performing a
fit of Eq.~(\ref{eqn:dln}) to the data.  We perform maximum likelihood
estimation~\cite{mood74} to determine the best estimate
parametrization of Eq.~(\ref{eqn:dln}); see Appendix~\ref{sect:mle} for
details.

After determining the parameters of the double log-normal
distribution, we conduct KS tests and Bayesian model selection as
before, and we find that a double log-normal has a posterior
probability of one when compared with the power-law model
(Fig.~\ref{ks_waiting}).  In fact, if we consider all three candidate
models simultaneously, we still find that the posterior probability of
the double log-normal is one (Fig.~\ref{threeway}).

\subsection{Alternative definition of the waiting times}

Recently, Barab\'asi and co-workers~\cite{vazquez05a,barabasi06} have
reinterpreted the definition of the waiting times introduced in
Ref.~\cite{barabasi05}.  Barab\'asi and co-workers note that the
actual waiting time should not be counted from the time the original
e-mail was sent, but from the time the original e-mail was first read.
This appears perfectly logical, but the database under investigation
does not provide us with information on when the user {\it actually\/}
first read the e-mail.  In fact, as we explained earlier the database
does not even provide information that would enable one to decide
whether an e-mail is a reply to a previous message or whether it is a
totally unrelated message.

Nonetheless, it is worthwhile to analyze in greater detail the manner
in which the authors of Refs.~\cite{vazquez05a,barabasi06} measure the
waiting time $\tau_w$ since they characterize it as an improvement
over the original method~\cite{barabasi05}.  At time $t_1$ user A
sends an e-mail to user B. At time $t_2>t_1$, user B sends an e-mail.
At time $t_3 \ge t_2$, user B sends an e-mail to user A.  The ``real''
waiting time is now defined as $\tau_r = t_3-t_2$, instead of $\tau_w
= t_3-t_1$.  Note that $t_2$ still is \emph{not} the actual time when
the user actually first read the e-mail.

\begin{figure}
\centerline{\includegraphics*[width=0.4\columnwidth]{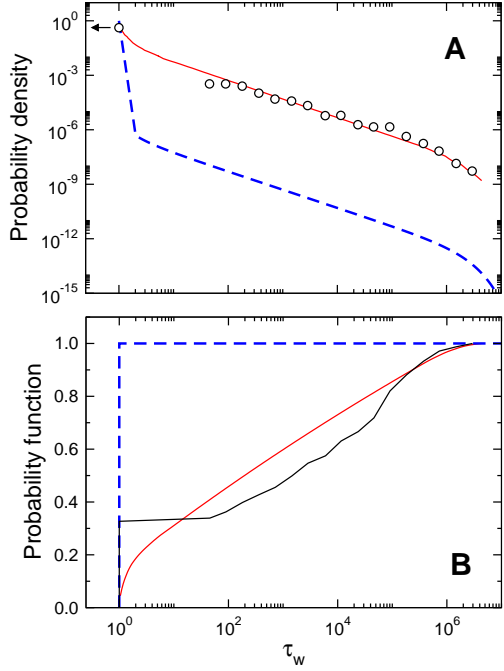}}
\vspace*{-0.3cm}
\renewcommand{\baselinestretch}{1.0}
\caption{Apparent agreement of new waiting time measure with priority
  queuing model prediction ($p=0.999999$,$L=2$) in Figs.~1a--b of
  Ref.~\cite{barabasi06}.  {\bf A}, Reproduction of the empirical
  probability density (open circles) and purported model solution (red
  line) from Fig.~1a of Ref.~\cite{barabasi06} using
  VistaMetrix~\cite{skillcrest04}.  To match the model with the
  empirical data, the authors of Ref.~\cite{barabasi06} claim that
  $\tau_r=0$ is actually $\tau_r=1$ (arrow).  Moreover, the authors of
  Ref.~\cite{barabasi06} do not use the actual model solution from
  Ref.~\cite{vazquez05} (blue dashed line) as claimed.  {\bf B},
  Probability function for the empirical waiting times, the purported
  model prediction, and the actual model solution.  Even if the
  purported model solution was correct, it is visually apparent that
  it does not match the large gap in waiting times between $\tau_w=1$
  and $\tau_w=60$ seconds.}
\label{reply_critique}
\end{figure}

We find three major problems with the reported predictive ability of
the priority queuing model to capture the peak, the power-law regime,
and the exponential cut-off of the waiting time distributions.  First,
we are troubled that the ``agreement'' for the peak at $\tau_r=1$ is
obtained by making the transformation $\tau_r=1$ if $t_3=t_2$, instead
of $\tau_r=0$ as would be expected from the definition.  In other
words, to match the peak at $\tau_r=1$, Barab\'asi and co-workers
state that $0=1$.

Moreover, we are surprised that Barab\'asi and co-workers claim to use
the exact model solution to predict the empirical waiting times.
Unlike Fig.~1a of Ref.~\cite{barabasi06}, the exact probability density
has a large, discontinuous drop at $\tau_r=1$~\cite{vazquez05}.  When
we compare the data presented in Ref.~\cite{barabasi06} with the actual
solution, it is clear that the model does not, in fact, match the
empirical data (Fig.~\ref{reply_critique}A).

Finally, there are no waiting time values for $\tau_w$ between 1 and
60 seconds whereas the priority queuing model predicts a smooth
continuous decrease of the probability density function in that
region.  While the difference between the two functions is difficult
to discern in the plot of Ref.~\cite{vazquez05a}, the difference is
actually quite marked (Fig.~\ref{reply_critique}B).

\section{The priority queuing model} \label{sect:pqm}

We also examined the priority queuing model presented to explain the
reported power-law in e-mail communication~\cite{barabasi05}.  This
model is defined as follows.  An individual has a priority queue with
$L$ tasks.  Each task is assigned a priority $x$ drawn from a uniform
distribution $\rho(x)=U[0,1]$.  At each unit time step, the user
executes either the highest-priority task with probability $p$ or a
randomly selected task with probability $1-p$. The executed task is
then removed from the queue and a new task with priority $x$, again
drawn from $\rho(x)$, is added to the queue.  For the sake of
comparison of the model predictions with the empirical data,
Barab\'asi surmised that a user's queue consists of e-mails which
require a response.  The model thus predicts the time $\tau_w$ that a
message spends in the user's inbox prior to response.

\begin{figure}
\centerline{\includegraphics*[width=0.5\columnwidth]{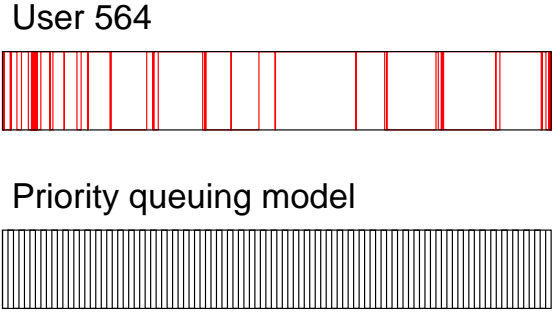}}
\vspace*{-0.3cm}
\renewcommand{\baselinestretch}{1.0}
\caption{Comparison of time series of a typical user versus the time
  series assumed in the priority queuing model.  The time series of
  100 activities for an actual user is quite different than the time
  series for 100 activities for the priority queuing model.  In the
  priority queuing model, one task is executed at each time step
  causing the interevent times to be distributed according to a
  Dirac-delta function.  As demonstrated in Section~\ref{sect:tau} and
  Ref.~\cite{barabasi05}, the interevent times are distributed with a
  heavy-tail.}
\label{modeltime}
\end{figure}

We first address the deficiencies in the model's assumptions.  First,
humans can only process a handful of pieces of information at any time
\cite{miller56}.  However, many users of e-mail hold tens,
hundreds, or even thousands of e-mails in their inbox which may require
action.  It is therefore unrealistic to expect any user to account for
each task's priority or to be able to carefully determine the absolute
(or even the relative) priority of such a large number of
tasks.

Secondly, the priority queuing model does not account for the
heterogeneities in interevent times revealed by our analysis in
Section~\ref{sect:tau} and reported in Ref.~\cite{barabasi05}.  In the
priority queuing model, tasks are executed at each time step which
means that the distribution of interevent times can be described by
Dirac's delta function (Fig.~\ref{modeltime}).

\begin{figure}
\centerline{\includegraphics*[width=0.7\columnwidth]{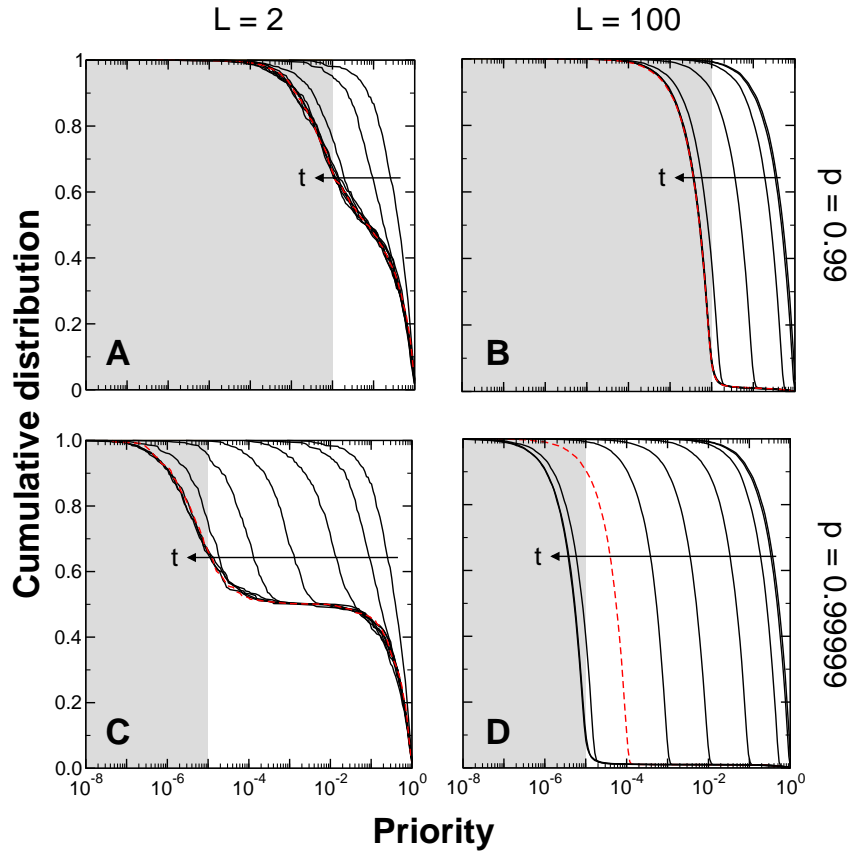}}
\vspace*{-0.3cm}
\renewcommand{\baselinestretch}{1.0}
\caption{Deducing the steady state behavior of the priority queuing
  model. {\bf A}--{\bf D} The cumulative distribution of priorities
  $x$ in queue after $10^i$ tasks have been executed where
  $i=0,1,2,\cdots,9$.  The author of Ref.~\cite{barabasi05}, considers
  the case of (D) $L=100$ and $p=0.99999$ after $10^6$ tasks have been
  executed (red dashed line).  In this case, however, the model has
  not even reached steady-state.  The important thing to notice is
  that after a short transient time, the priorities become uniformly
  distributed in the very small interval $[0,1-p]$ denoted with the
  gray shading.  When new tasks are added to the queue from a uniform
  distribution on the interval $[0,1]$, the vast majority of new tasks
  are executed immediately.  This feature of the priority queuing
  model is not representative of human behavior.  }
\label{x_evo}
\end{figure}

The priority queuing model also suffers from several unrealistic
predictions.  First, the time for the model to reach steady-state
increases as $L/(1-p)$ \cite{vazquez05}.  This means that for the case
considered in Ref.~\cite{barabasi05} ($L=100$, $p=0.99999$), the time
to reach steady-state is on the order of $10^7$ tasks.  If a user
operating according to those parameter values sends 100 e-mails a day
(a very large number of e-mails), it would take him 100,000 days
$\approx$ 300 years to reach steady-state.  It is also worthwhile to
note that the results for the model in Ref.~\cite{barabasi05} were
not even obtained for steady-state, implying that the data is actually
a mixture of different stages of the relaxation process of the model.

\begin{figure}
\centerline{\includegraphics*[width=0.8\columnwidth]{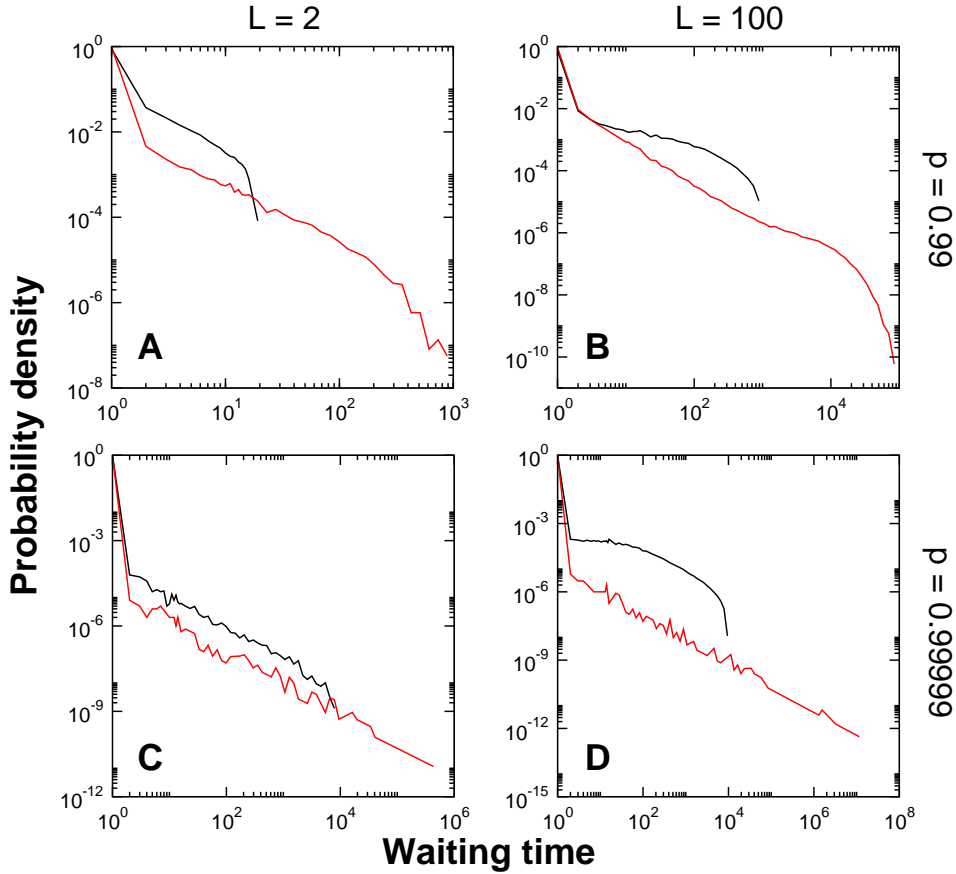}}
\vspace*{-0.3cm}
\renewcommand{\baselinestretch}{1.0}
\caption{Contrasting the transient and steady-state behavior of the
  priority queuing model. {\bf A}--{\bf D}, We plot the distribution of waiting
  times during the transient period for times $t<0.1~L/(1-p)$
  (red) and steady-state for times $t>10~L/(1-p)$ (black).
  Notice that the likelihood that a task is executed after spending
  more than one unit of time on the queue is vanishingly small.  }
\label{trans_ss_tauw}
\end{figure}

Second, after reaching steady-state, the dynamics of the model become
quite anomalous. The priorities of the tasks in the user's queue
converge to a uniform distribution $U[0, 1-p]$ (Fig.~\ref{x_evo}),
while new tasks arrive with a priority drawn from $U[0,1]$.  Thus, in
the limit $p \rightarrow 1$, an e-mail user has a queue consisting of
extremely low priority tasks and consequently performs new tasks with
probability $p \rightarrow 1$ immediately upon arrival.  This results
in a peak at $\tau_w = 1$ that accounts for nearly all of the
probability mass (Fig.~\ref{trans_ss_tauw}).  Clearly this situation
is not representative of e-mail activity, let alone human behavior as
Ref.~\cite{barabasi05} claims.

More fundamentally, the priority queuing model predicts a distribution
of waiting times that decays as a power-law with an exponent
$\alpha=1$~\cite{vazquez05,vazquez05a} whereas the actual data clearly
rejects that description; cf.~Section~\ref{sect:tauw}.  In fact, a
superposition of two log-normals, one corresponding to waiting times
of less than a day and another corresponding to a waiting time of
several days provides an excellent description of the empirical data.
More importantly, that description agrees with the common experience
of e-mails users: one replies to e-mails within the day, if not,
within the next few days, and if not then, never.

\section{Conclusions} \label{sect:concl}

Here, we have quantitatively analyzed human e-mail communication
patterns.  In particular, we have found that the interevent times are
well-described by a log-normal distribution while the waiting times
are well-described by the superposition of two log-normal
distributions.  We have simultaneously rejected the hypothesis that
either quantity is adequately described by a truncated power-law with
exponent $\alpha \simeq 1$.  

We have also critically examined the priority queuing model proposed
by Barab\'asi to match the empirically observed waiting time
distributions.  After detailed analysis, we conclude that neither the
assumptions nor the predictions of the model are plausible.  We note
that the model does not match the empirically observed waiting time
distribution, and we therefore contend that the theoretical
description of human dynamics is an open problem.

Barab\'asi and coworkers have also examined the dynamics of letter
writing~\cite{oliveira05}, web browsing, library loans, and stock
broker transactions~\cite{vazquez05a}.  They argue that these
processes also follow power-law distributions and are consequences of
similar priority queuing processes.  Our analysis demonstrates that
care must be taken when describing data with fat-tails, particularly
when the apparent scaling exponent is close to one and the probability
distribution is concave.

\appendix

\noindent \section{Maximum likelihood estimation}
\label{sect:mle}

In maximum likelihood estimation, the likelihood function
$\mathcal{L}$ for distribution model $\mathcal{M}_j$ given the data
$\left\{ u_{w,i} \right\}$ is
\begin{equation}
  \mathcal{L}\left(\mathcal{M}_j | \left\{ u_{w,i} \right\} \right) =
  \prod_{k=1}^{N} p \left( u_{w,k} | \mathcal{M}_j \right)\,,
\end{equation}
where $N$ is the number of data points in the sample, the $\left\{
u_{w,i} \right\}$ are the empirical data points and $p(u_w |
\mathcal{M}_j)$ is the probability density function for the candidate
model $\mathcal{M}_j$ evaluated at each empirical data point.  We then
maximize the likelihood $\mathcal{L}$ to find the parametrization of
the model distribution $p(u_w | \mathcal{M}_j)$ that best approximates
the data.  For the double log-normal model, $p(u_w |$double
log-normal$) = p(u_w;f,\mu_1,\sigma_1,\mu_2,\sigma_2)$.  To find the
best estimate of the five parameters
$\{f,\mu_{1},\sigma_{1},\mu_{2},\sigma_{2}\}$, we obtain preliminary
estimates for $\mu_{1}$ and $\sigma_{1}$ from the mean and standard
deviation of $\left\{ u_{w,i} \right\}$ and subsequently maximize
likelihood $\mathcal{L}$ to find the appropriate double log-normal
parameters.  In practice, however, one typically performs a
minimization of $\mathcal{H} = -\ln
\mathcal{L}$~\cite{mood74,press02}.


\end{document}